\documentclass[%
 reprint,
 superscriptaddress,
 amsmath,amssymb,
 aps,
 prl,
 floatfix,
]{revtex4-1}
\usepackage{float}
\usepackage{graphicx}
\usepackage{dcolumn}
\usepackage{bm}
\usepackage{xcolor}

\usepackage{soul}
\usepackage{comment}

\begin{document}


\title{Experimental search for invisible dark matter axions around 22\,$\mu$eV}

\author{Younggeun Kim}
\thanks{These authors contributed equally to this work.}
\affiliation{Center for Axion and Precision Physics Research, IBS, Daejeon 34051, Republic of Korea}
\author{Junu Jeong}
\thanks{These authors contributed equally to this work.}
\affiliation{Center for Axion and Precision Physics Research, IBS, Daejeon 34051, Republic of Korea}
\author{SungWoo Youn}
\email{swyoun@ibs.re.kr}
\affiliation{Center for Axion and Precision Physics Research, IBS, Daejeon 34051, Republic of Korea}%
\author{Sungjae Bae}
\affiliation{Department of Physics, KAIST, Daejeon 34141, Republic of Korea}
\affiliation{Center for Axion and Precision Physics Research, IBS, Daejeon 34051, Republic of Korea}
\author{Kiwoong Lee}
\affiliation{Center for Axion and Precision Physics Research, IBS, Daejeon 34051, Republic of Korea}
\author{Arjan F. van Loo}
\affiliation{RIKEN Center for Quantum Computing (RQC), Wako, Saitama 351-0198, Japan}
\affiliation{Department of Applied Physics, Graduate School of Engineering, The University of Tokyo, Bunkyo-ku, Tokyo 113-8656, Japan}
\author{Yasunobu Nakamura}
\affiliation{RIKEN Center for Quantum Computing (RQC), Wako, Saitama 351-0198, Japan}
\affiliation{Department of Applied Physics, Graduate School of Engineering, The University of Tokyo, Bunkyo-ku, Tokyo 113-8656, Japan}
\author{Seonjeong Oh}
\affiliation{Center for Axion and Precision Physics Research, IBS, Daejeon 34051, Republic of Korea}
\author{Taehyeon Seong}
\affiliation{Center for Axion and Precision Physics Research, IBS, Daejeon 34051, Republic of Korea}
\author{Sergey Uchaikin}
\affiliation{Center for Axion and Precision Physics Research, IBS, Daejeon 34051, Republic of Korea}
\author{Jihn E. Kim}
\affiliation{Department of Physics, Seoul National University, Seoul 08826, Republic of Korea}
\author{Yannis K. Semertzidis}
\affiliation{Center for Axion and Precision Physics Research, IBS, Daejeon 34051, Republic of Korea}
\affiliation{Department of Physics, KAIST, Daejeon 34141, Republic of Korea}

\date{\today}

\begin{abstract}
The axion has emerged as the most attractive solution to two fundamental questions in modern physics related to the charge-parity invariance in strong interactions and the invisible matter component of our universe.
Over the past decade, there have been many theoretical efforts to constrain the axion mass based on various cosmological assumptions.
Interestingly, different approaches from independent groups produce good overlap between 20 and 30\,$\mu$eV.
We performed an experimental search to probe the presence of dark matter axions within this particular mass region.
The experiment utilized a multi-cell cavity haloscope embedded in a 12\,T magnetic field to seek for microwave signals induced by the axion--photon coupling.
The results ruled out the KSVZ axions as dark matter over a mass range between 21.86 and 22.00\,$\mu$eV at a 90\% confidence level.
This represents a sensitive experimental search guided by specific theoretical predictions.
\end{abstract}

\maketitle

The absence of charge-parity (CP) violation in quantum chromodynamics (QCD), known as the strong-CP problem, is a long-standing enigma in particle physics.
One of the most favored solutions, called the Peccei-Quinn (PQ) mechanism~\cite{PecceiQuinn:PRL:1977}, involves the spontaneous breaking of a new global symmetry that results in the emergence of a pseudo-Nambu–Goldstone boson, the axion ~\cite{Weinberg:PRL:1978, Wilczek:PRL:1978}.
The experimental exclusion of the ``visible" axion model, PQWW (Peccei-Quinn-Weinberg-Wilczek), has led to two classes of ``invisible" models, KSVZ (Kim-Shifman-Vainshtein-Zakharov)~\cite{Kim:PRL:1979, Shiftman:NPB:1980} and DFSZ (Dine-Fischler-Srednicki-Zhitnitsky)~\cite{Zhitnitsky:SJNP:1980, Dine:PLB:1981}.
This feebly interacting particle raised further interest due to its cosmological role as dark matter~\cite{Preskill:PLB:1983, Abbott:PLB:1983, Dine:PLB:1983}, a mysterious substance believed to constitute $\sim$85\% of the matter in the universe. 
Various experimental efforts have been made and are currently underway to find evidence for axions in our galactic halo~\cite{cajohare:github:2020, Yannis:SciAdv:2022}.

A common method to search for axions relies on electromagnetic interactions provoked by strong magnetic fields.
In particular, the cavity haloscope takes advantage of the resonant enhancement of photon signals~\cite{Sikivie:PRL:1983}, providing the most sensitive method in the microwave region.
The expected power of photons converted from axions is given by~\cite{Sikivie:PRD:1985, Kim:JCAP:2020}
\begin{equation}
    P_{\mathrm{sig}} = g_{a\gamma\gamma}^2\frac{\rho_a}{m_a}\langle \mathbf{B}_{e}^{2} \rangle V_{c}G\frac{Q_{l}Q_{a}}{Q_{l} + Q_{a}}\frac{\beta}{1+\beta},
    \label{eq:conv_power}
\end{equation}
where $g_{a\gamma\gamma}$ is the axion--photon coupling, $\rho_{a}$ and $m_a(\simeq 2\pi\nu_a)$ are the local density and mass (Compton frequency) of the dark matter axion, $\langle \mathbf{B}_{e}^{2} \rangle$ is the average of the square of the external magnetic field inside the cavity volume $V_c$, $G$ is the geometry factor of the resonant mode under consideration, $Q_l(=Q_c/(1+\beta))$ and $Q_a$ are the loaded (unloaded) cavity and axion quality factors, and $\beta$ is the antenna coupling coefficient.
The experimental performance is represented by the signal-to-noise ratio (SNR) 
\begin{equation}
    {\rm SNR} \equiv \frac{P_{\rm sig}}{\delta P_{\rm sys}} = \frac{P_{\rm sig}}{k_B T_{\rm sys}}\sqrt{\frac{\Delta t}{\Delta \nu}},
    \label{eq:snr}
\end{equation}
where $\delta P_{\rm sys}=P_{\rm sys}/\sqrt{\Delta \nu \Delta t}$ describes fluctuations in system noise power over a bandwidth $\Delta \nu$ during an integration time $\Delta t$~\cite{Dicke:RSI:1946}, and $T_{\rm sys}$ is the equivalent system noise temperature transformed using $P_{\rm sys}=k_B T_{\rm sys}\Delta\nu$~\cite{Johnson:PR:1928, Nyquist:PR:1928}.
The system noise is given by a linear combination of the thermal noise from the cavity and the added noise from a receiver chain.

Over the past decade, numerous theoretical attempts have been made to constrain the mass of dark matter axion, depending on its formation mechanism and methodological approach for its evolution~\cite{Wantz:PRD:2010, Kawasaki:PRD:2015, Borsanyi:Nature:2016, Bonati:JHEP:2016, Klaer:JCAP:2017, Dine:PRD:2017, Buschmann:PRL:2020, Buschmann:NatureComm:2022}.
It is noteworthy that in the scenario where the PQ symmetry is broken after cosmic inflation, the mass ranges predicted by several independent theory groups share a common region between 20 and 30\,$\mu$eV~\cite{Wantz:PRD:2010, Borsanyi:Nature:2016, Klaer:JCAP:2017, Buschmann:PRL:2020}.
We, at the Center for Axion and Precision Physics Research (CAPP), therefore set up a dedicated experiment to search for dark matter axions as a test of post-inflationary axion cosmology in this specific mass region.
In this Letter, we report the results of the cavity haloscope search between 21.86 and 22.00\,$\mu$eV with theoretically interesting sensitivity.

\begin{figure}[b]
\centering
\includegraphics[width=0.95\linewidth]{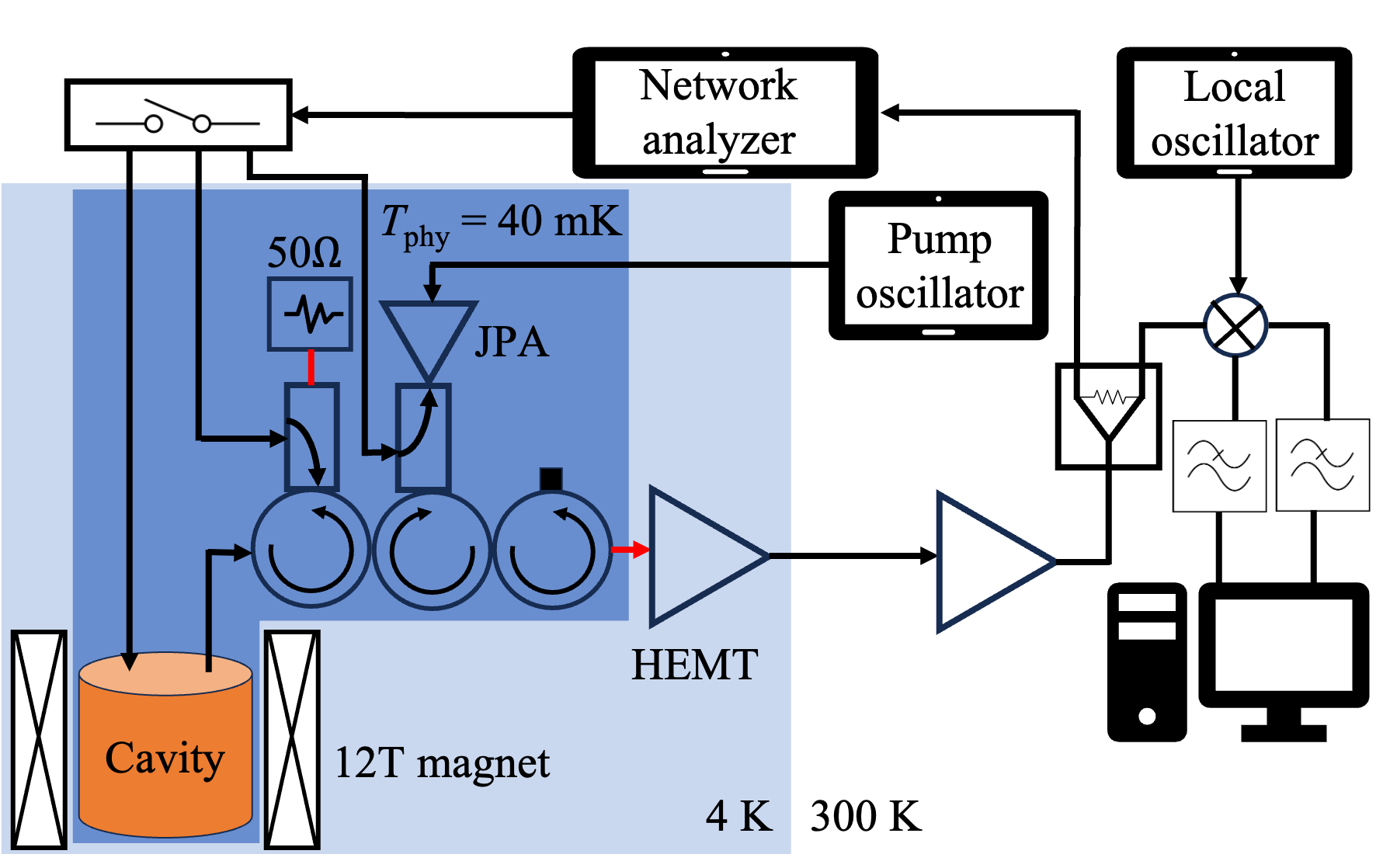}
\caption{Schematic of the CAPP-12T haloscope setup.}
\label{fig:exp_schematic}
\end{figure}

The major equipment of the haloscope is a 12-T, $\O$96-mm superconducting solenoid and a cryogenic dilution refrigerator (DR).
The solenoid operating in persistent mode provides an average field of 9.8\,T within the cavity peaking at 12\,T at the magnet center, and the DR is capable of maintaining the detector components attached to its mixing chamber below 40\,mK.
The cryogenic components include a microwave cavity and a readout chain, which consists of a Josephson Parametric Amplifier (JPA) and a pair of High Electron Mobility Transistor (HEMT) amplifiers.
The axion signal is further amplified at room temperature, translated to an intermediate frequency (IF), converted to a digital format, and then transformed into the frequency domain before storage.
The haloscope was named ``CAPP-12T" after the strength of the magnetic field and the experimental setup is schemed in Fig.~\ref{fig:exp_schematic}.

\begin{figure}
\centering
\includegraphics[width=0.8\linewidth]{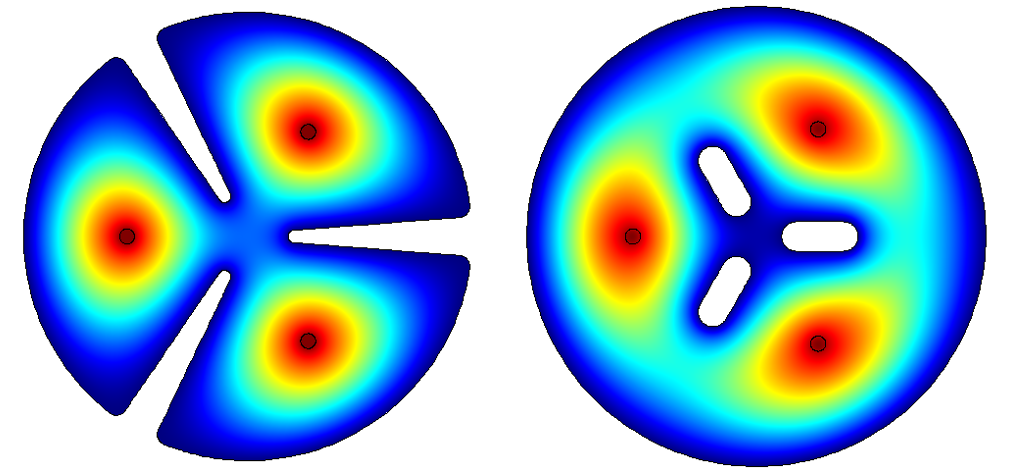}
\caption{Electric field distribution of 3-cell cavities for the original (left) and modified (right) design, with dielectric rods (small black circles) located in the center of each cell.
}
\label{fig:MC cavity}
\end{figure}

The resonant cavity employed a multi-cell design~\cite{JEONG2018412} to effectively increase the search frequency while utilizing the available magnet volume to its fullest extent.
The original design was modified with metallic partitions detached from the single-body structure and placed separately within the cavity, as seen in Fig.~\ref{fig:MC cavity}.
This design reduces the complexity of cavity machining and thus improves fabrication and assembly precision.
Moreover, the additional space between the partitions and cavity wall allows the individual cells to couple more strongly, naturally alleviating field localization.
A simulation study showed that, despite a slight reduction in cavity quality factor, such a configuration enhances the geometry factor and thus improves the overall performance of haloscope search~\cite{10.3389/fphy.2024.1347003}.
A 3-cell design was chosen for the resonant frequency of our desired mode (TM$_{010}$-like) to sit between 5 and 6\,GHz while making full use of the given magnet volume.
The cavity made of oxygen-free copper with internal dimensions of $\O 78\,{\rm mm} \times 300\,{\rm mm}$ has a detection volume of 1.38\,L.
Similar to the tuning mechanism described in Ref.~\cite{Junu:PRL:2020}, a set of alumina rods ($\O 2.6\,\rm{mm}$) and a single rotary actuator were used to tune the frequency between 5.20 and 5.35\,GHz.
The typical parameter values in this range are $Q_{c}=3.9\times10^{4}$ and $G = 0.68$.
A pair of dipole antennas were introduced: one is strongly coupled to capture the signal with $\beta\sim2$ and the other is weakly coupled to allow transmission measurements and synthetic signal injection.
Assuming the KSVZ axion constitutes the entire dark matter, i.e., $\rho_{a} = 0.45\,\rm{GeV/cm^{3}}$, the expected signal power is $3.8\times10^{-23}$\,W.

JPAs are featured by noise near the quantum limit~\cite{Zhong_2013} and thus play a crucial role in increasing the sensitivity of experiments.
Used as the first stage amplifier in the receiver chain, our flux-driven JPA is of the same type as the one characterized in Ref.~\cite{Kutlu_2021}.
The JPA was protected from external magnetic fields up to 0.1\,T by a three-layer shield of aluminum, cryoperm and NbTi alloys (inside to outside)~\cite{Sergey:LT29:2023}.
A series of RF circulators and an isolator were configured in a manner to reduce interference between the cavity, JPA, and HEMTs.
A directional coupler connected to the first circulator provides a dual route for cavity reflection measurement and noise figure evaluation.
A pair of HEMTs, thermally anchored to the 4K plate, constitute the cryogenic receiver chain.
The noise temperature of the HEMT chain was estimated based on the Y-factor method using a 50-$\Omega$ terminator with a heater directly attached, resulting in 3.2--3.5\,K, including the contribution from the upstream components, over a wider frequency range.
The noise source was connected to the directional coupler via a superconducting line shown as a red line in Fig.~\ref{fig:exp_schematic}.
At room temperature, additional amplification yielded a total gain exceeding 95\,dB.
The signal was down-converted to 3\,MHz using an IQ mixer where the in-phase and quadrature components are processed separately until digitized at a sampling rate of 20\,MHz and merged in software.
This time series data was Fourier-transformed into the frequency domain with a resolution bandwidth of 100\,Hz within a 1-MHz span and stored on tape for post-analysis.

\begin{figure}
\centering
\includegraphics[width=.8\linewidth]{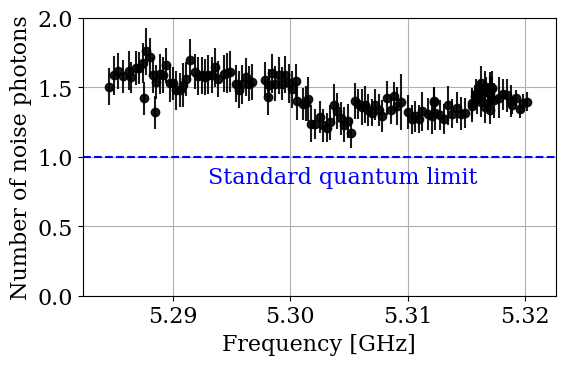}
\caption{System noise measured in photon number around 5.3\,GHz.}
\label{fig:sys_noise}
\end{figure}

The noise performance of the receiver chain was characterized by the Noise Visibility Ratio (NVR), defined as the excess noise visible in the power spectrum when the pump is turned on versus off~\cite{Friis:ProcIRE:1944, PhysRevX.5.041020}.
This technique allows for time-efficient {\it in-situ} measurement of noise temperature.
Our JPA exhibited noise close to the standard quantum limit over its tunable range of 5.10 to 5.35\,GHz, with a typical gain of 22\,dB.
Combined with the cavity physical temperature of $\sim$40\,mK, which yields an effective temperature, $T_{\mathrm{eff}} = (h\nu/k_B)[1/(e^{h\nu/k_BT_{\rm phy} } - 1) + 1/2 ]$, of 130\,mK around 5.3\,GHz, the system noise was estimated to be $T_{\rm sys} = 360-410$\,mK. 
This corresponds to approximately 1.5 noise photons as shown in Fig.~\ref{fig:sys_noise}.
The methodological validity was confirmed by comparing with the Y-factor method performed using the noise source, which showed agreement within 2\%.
The effect of impedance mismatch, estimated from the residual structure of the spectrum after dividing the JPA gain, was also taken into account.

Data acquisition (DAQ) begins with the determination of cavity parameters using a network analyser (NA): $Q_c$ via transmission measurement and $\beta$ by fitting the circles on the Smith chart~\cite{IEEE:Q_coupling:1984}.
The JPA resonance was set about 200\,kHz away from the cavity resonance in order to operate in the phase-insensitive mode.
The Nelder-Mead (NM) algorithm~\cite{NelderMead:CJ:1965} was adopted as a heuristic search method for the optimal flux and pump that minimize the noise temperature for a given resonant frequency.
The test points of the initial simplex were given based on the resonance and gain profiles obtained by sweeping the flux and pump power, respectively, prior to experimentation.
At each iteration over the two-dimensional space, the search algorithm found a new test point.
The test point was evaluated in terms of power saturation of the JPA using the NA.
If an increase in JPA gain of $>$10\% was observed for 3-dB weaker input power, another new test point with 0.01-dB lower pump power was set and re-evaluated.
Once the requirement was met, the added noise was measured using the NVR technique described earlier.
The number of iterations was set to 10, which was sufficient to converge to the point that yields the lowest noise temperature, taking up to 3 minutes.
The NM algorithm provides an \textit{in-situ} method to characterize the JPA, thereby substantially increasing the reliability of measurements compared to a predetermined set of parameters obtained during the commissioning phase.

The minimum size of data that was averaged and stored corresponds to 10 seconds.
DAQ was interrupted every 6 minutes to examine the stability of JPA operation.
A deviation in gain exceeding 1\,dB from the prior measurement signifies the need of establishing a new JPA operating point, and the NM algorithm was repeated.
To reach KSVZ sensitivity, data was collected for 7--8 hours at each tuning step depending on the JPA performance with a DAQ efficiency, defined as $(t_{\rm total}-t_{\rm dead})/t_{\rm total}$, of 91.3\%.
The search frequency was then tuned by 150\,kHz, which is less than half the loaded cavity bandwidth.
The data set for the results reported in this Letter was collected over 82 days, from July 1st to November 15th, 2023, with several weeks of downtime due to system maintenance.

The data analysis follows the procedure described in Ref.~\cite{Brubaker:PRD:2017}.
For each 10-s raw spectrum, the power in the individual bins was normalized by the baseline power, obtained using the Savitzky-Golay (SG) filter~\cite{Savitzky1964}, and subtracting unity from it.
The power excess was then scaled by multiplying by $(P_{\rm sig}/\delta P_{\rm sys})^{-1}$, where the KSVZ axions were assumed in $P_{\rm sig}$, so that the resulting power excess represents the expected SNR in each bin.
A series of scaled spectra obtained over consecutive tuning steps were combined bin by bin to construct a broad spectrum spanning the entire search range.
Finally, this vertically combined spectrum was further processed by merging the power of the frequency bins, weighted by the boosted Maxwell-Boltzmann probability density function (MB PDF), which describes the standard halo axion distribution~\cite{Turner:PRD:1990}.
This grand spectrum represents the maximum likelihood estimate of the axion model assuming the axion mass lies in each bin.
Frequency bins with high excess power become candidates for rescanning.

The fitting parameters of the SG filter were determined based on an extensive Monte Carlo (MC) study to maximize the SNR efficiency, $\epsilon_{\rm SNR}$, in the presence of signal.
We generated axion signals with an excess of 5 on top of the raw spectra and repeated the entire analysis procedure for various polynomial degrees and window sizes.
We found that a degree of 4 and a size of 1101 bins yielded a signal efficiency of $84.7\pm3.1\%$ and a noise spectrum following a zero-centered normal distribution with a standard deviation of 0.91, denoted as $\mathcal{N}(0.00,0.91)$, resulting in $\epsilon_{\rm SNR}=93.0\pm3.4\%$.
The width of the grand spectrum was re-normalized to unity for the convenience of utilizing the standard normal distribution $\mathcal{N}(0,1)$.

\begin{figure}
\centering
\includegraphics[width=.9\linewidth]{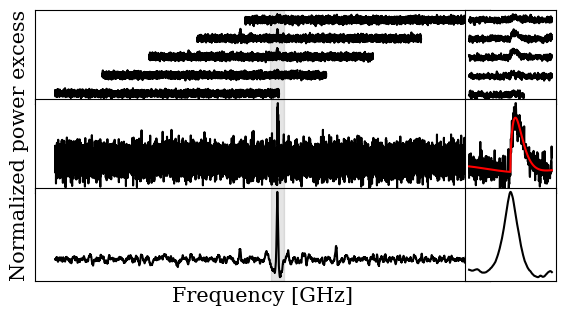}
\caption{\label{fig:synthetic_axion}(Top) A series of normalized power spectra containing the synthetic axion signal. 
(Middle) Vertically combined spectrum fitted with the MB signal PDF on a 4th-order polynomial background (red line). 
(Bottom) Corresponding grand spectrum.}
\end{figure}

To verify the performance of the detection system and data analysis, we conducted a blind analysis for a synthetic axion signal injected into the cavity at unknown frequency.
The injected power was calibrated by examining the SNR of the output through the chain with the JPA off, for which the noise level was well measured using the Y-factor technique.
The axion line shape was implemented by generating monochromatic signals every 10\,ms with equal power in frequency bins randomly selected according to the boosted MB PDF.
Through our data analysis, the signal was successfully reconstructed, as shown in Fig.~\ref{fig:synthetic_axion}. 
The vertical spectrum was fitted using the input signal function to return $\nu_a = 5316.3140 \pm 0.0001$\,MHz and $Q_a = (1.12\pm0.08)\times10^6$, consistent with the values of the injected signal of 5316.3140\,MHz and $1.09\times10^6$, respectively.
Using these parameters, the axion-photon coupling was also evaluated to be $g_{a\gamma\gamma}/g_{a\gamma\gamma}^{\rm KSVZ} = 4.97\pm0.04$, consistent with both the injected value of $5.00\pm 0.04$ and the peak value directly read from the grand spectrum, $4.93$.
After identifying the blind signal injection, additional data was collected in this frequency region.

A null hypothesis concerning axions with ${\rm SNR}=5$ expects a distribution $\mathcal{N}(5.0,1.0)$.
The threshold for a 90\% confidence level (CL) was set at $3.718\sigma$.
Out of a total of 367,215 frequency bins across the grand spectrum, 65 bins exceeding the threshold were clustered into six candidates.
For each candidate, an additional hour of data was collected and statistically added to the original data set to be re-analyzed.
Up to four iterations of this re-scanning process attenuated all the candidates below the threshold, confirming they were attributed to statistical fluctuations.

\begin{figure*}
\includegraphics[width=0.9\linewidth]{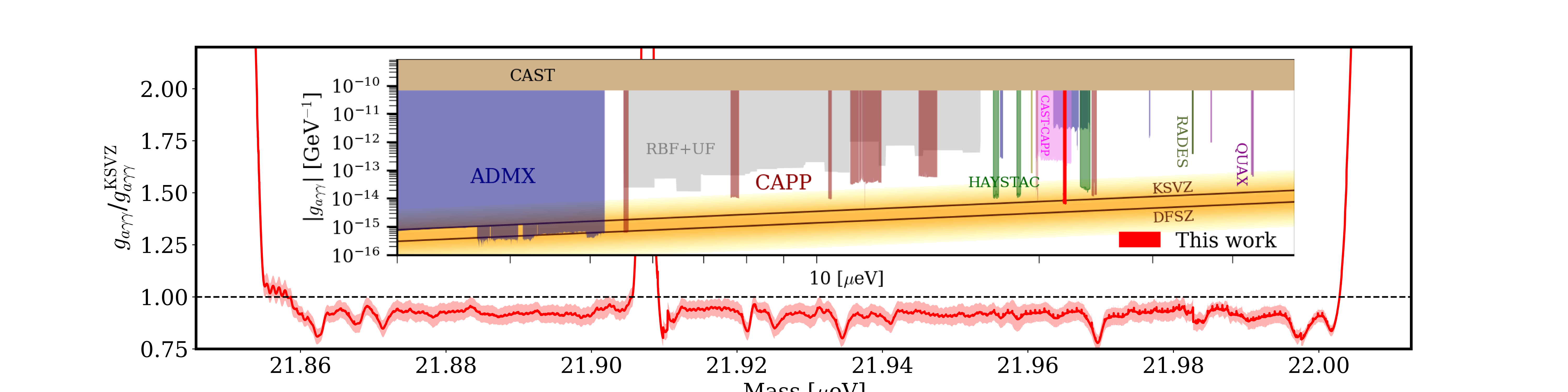}
\caption{\label{fig:exclusion}Exclusion limits on axion--photon coupling set by CAPP-12T at 90\% CL. 
The light red band represents the total uncertainty of measurements described in the text.
The mode causing the mixing observed between 21.90 and 21.92\,$\mu$eV was identified as the TE$_{313}$ mode. 
The results from other experiments~\cite{PhysRevLett.59.839,PhysRevD.42.1297,PhysRevLett.80.2043,PhysRevD.64.092003,PhysRevLett.118.061302,PhysRevLett.120.151301,PhysRevLett.124.101303,PhysRevLett.124.101802,Junu:PRL:2020,Alvarez-Melcon:2021aa,Backes:2021aa,PhysRevLett.126.191802,PhysRevLett.127.261803,Adair:2022aa,Kutlu:2022kvo,PhysRevLett.128.241805,PhysRevD.106.052007,PhysRevLett.129.111802,PhysRevLett.130.071002,Anastassopoulos:2017aa} are also shown in the inset.
}
\end{figure*}

With no signal observed, exclusion limits were set on axion--photon coupling under the null hypothesis.
Since the power excess was normalized to the KSVZ axion signal during the scaling process, the reciprocal of the standard deviation $\sigma$ represents the final SNR in each bin of the entire spectrum.
Taking the SNR efficiency into account, the axion--photon coupling is given by $\sqrt{5\sigma/\epsilon_{\rm SNR}}$ in units of KSVZ coupling.
Finding an average $\sigma=0.16$ with no significant excess in our data, we excluded dark matter axions with coupling $g_{a\gamma\gamma} \gtrsim  0.93\times g_{a\gamma\gamma}^{\rm KSVZ}$ at 90\% CL in the mass range between 21.86 and 22.00\,$\mu$eV.
Figure~\ref{fig:exclusion} shows our exclusion limits along with other experimental results.

Various uncertainties in setting the exclusion limits were assessed.
The largest uncertainty occurred in noise temperature measurements.
The statistical contribution of 30\,mK was estimated from the fluctuations in noise measured every 6 minutes at a given frequency and the systematic contribution was obtained from the difference between two independent (NVR and Y-factor) methods.
These resulted in an uncertainty of 8.1\% in noise estimation.
The uncertainty quoted for $\epsilon_{\rm SNR}$ was the second largest contributor.
Besides, Smith circle fitting for estimating the antenna coupling returned errors of up to 1.7\%.
Statistical fluctuations in loaded $Q$ measurements read a typical value of 220, giving a relative uncertainty of 1.3\%. 
Finally, a quadratic sum of these individual uncertainties yielded a total uncertainty of 4.7\% in determining $g_{a\gamma\gamma}$, which is visualized in Fig.~\ref{fig:exclusion} as the light red band.

In summary, we performed an experimental search near 22\,$\mu$eV for the invisible QCD axion as a favored dark matter candidate appearing in the post-inflationary scenario.
The experiment featured a modified multi-cell cavity immersed in a 12-T magnetic field and a JPA whose characteristics were determined {\it in-situ} using the Nelder-Mead algorithm.
The search relied on the axion--photon coupling and the null results ruled out the KSVZ axion as dark matter in the mass range 21.86$-$22.00\,$\mu$eV with the most stringent limits set to date.
This corresponds to an experimental effort to probe a specific mass region guided by particular theoretical predictions with significant sensitivity.
Experimental efforts will continue by extending sensitive searches to explore as much of the parameter space preferred by theoretical calculations as possible.

\begin{acknowledgments}
This work was supported by the Institute for Basic Science (IBS-R017-D1) and JSPS KAKENHI (Grant No.~JP22H04937). 
A. F. Loo was supported by a JSPS postdoctoral fellowship and J. E. Kim was partially supported by Korea National Science Foundation.
\end{acknowledgments}

\bibliographystyle{apsrev4-2}
\bibliography{main}

\end{document}